\documentclass{article}

\usepackage{amsfonts}
\usepackage[dvips]{graphicx}
\usepackage{epsfig}
\usepackage{color}
\usepackage{float}
\title{Local orientational order in the Stockmayer liquid}
\author{V.N. Blinov\footnote{blinov.veniamin@gmail.com}\\
Department of mechanics and mathematics\\
The Lomonosov Moscow State University}

\begin{document}
\maketitle

\begin{abstract}
Phase behaviour of the Stockmayer fluid is studied with a method similar to the Monte-Carlo annealing scheme. We introduce a novel order parameter which is sensitive to the local co-orientation of the dipoles of particles in the fluid. We exhibit a phase diagram based on the behaviour of the order parameter in the density region $0.1 \leq \rho^* \leq 0.32$. Specifically, we observe and analyse a second order \textit{locally disordered fluid $\rightarrow$ locally oriented fluid} phase transition.
\end{abstract}

\section{INTRODUCTION}
The study of dipolar spheres models is of significant importance as they can be considered as native models on ferrofluids \cite{rosensweig} or dispersions of polar molecules, or as simple media with a property of self-organization, \cite{douglas}.\\
\\
There are three models frequently studied in the context of dipolar fluids: the dipolar hard sphere (DHS) model, the dipolar soft sphere (DSS) model, and the Stockmayer (SM) model. The total potential energy of each of these models has the following form:
$$
U_{tot} = \sum_{i \neq j}\left( U_{dd}(i, j) + U_{sr}(i, j)\right)
$$ 
with $U_{dd}$ being a dipole-dipole potential, $U_{sr}$ -- a short range potential and summation being performed over all pairs of particles. \\
The first term $U_{dd}$ is common for all three models; they differ with the respect to their short-range potentials:\\
\noindent
the DHS model has the hard sphere short range potential \\
$$
U_{hs}^{ij} = \left\{
\begin{array}{rcl}
 0,      &  \mbox{if} & r_{ij} \geq R, \\
 \infty, &  \mbox{if} & r_{ij} < R.
\end{array} 
\right.
$$
the DSS model employs the soft sphere potential \\
$$
U_{ss}^{ij} = 4 \epsilon \left(\frac{\sigma}{r_{ij}}\right)^{12}\\
$$
the Stockmayer (SM) model is distinguished by its use of the Lennard-Jones short range potential \\
$$
U_{lj}^{ij} = 4 \epsilon\left[\left(\frac{2R}{r_{ij}}\right)^{12}-\left(\frac{2R}{r_{ij}}\right)^6\right].
$$
In these formulae, $r_{ij}$ stands for the distance between i-th and j-th particle. \\
In the current paper we shall pay attention to the last model mentioned. \\
The potential energy of the Stockmayer fluid of N particles is written as
$$
U = \sum_{1 \leq i < j \leq N} \left[ U_{lj}^{ij}+U_{dd}^{ij} \right]
$$
with
$$
U_{lj}^{ij} = 4 \epsilon\left[\left(\frac{2R}{|\vec{r}_i-\vec{r}_j|}\right)^{12}-\left(\frac{2R}{|\vec{r}_i-\vec{r}_j|}\right)^6\right]
$$
and
$$
U_{dd}^{ij} = \frac{\vec{D}_i\cdot\vec{D}_j}{|\vec{r}_i-\vec{r}_j|^3} -
3 \frac{(\vec{D}_i\cdot(\vec{r}_i-\vec{r}_j))(\vec{D}_j\cdot(\vec{r}_i-\vec{r}_j))}{|\vec{r}_i-\vec{r}_j|^5}.
$$
\\

\begin{figure}[H]
\center{
\includegraphics[width=5.5cm]{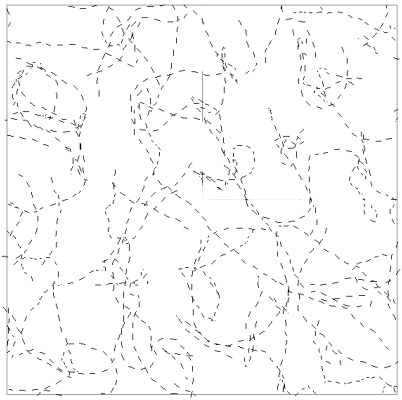}
\caption{String phase in the Stockmayer fluid at low densities. Projections of dipoles on a plane are shown.}
\label{fig:chains}
}
\end{figure}

Study of the DS models uses numerical methods since no theory for quantitative description has been developed. The Monte-Carlo and molecular-dynamical studies resulted in significant achievements in discovering the phase behaviour of these models. A large part of these results is summarized in a review by Weis and Levesque, \cite{weis_levesque}. \\
\\
Thus, previous studies of this topic have shown that all three basic models form an isotropic phase at high temperatures. At the same time, at high densities these models demonstrate low-temperature crystal phases, \cite{crystal, crystal2}. This difference of phases arouses interest in the phase transitions in these systems.\\
\\
Recent studies of phase transitions have revealed the presence of the \textit{isotropic liquid $\rightarrow$ orientationally ordered liquid} phase transition in dilute DS fluids. Specifically, at low densities the SM fluid exhibits the so-called \textit{string liquid} phase, formed by chains of dipolar particles with dipoles co-orientated in each chain, see fig. \ref{fig:chains}.\\
\\
Despite all the similarities of the three models, the Stockmayer model demonstrates more complicated behaviour due to the Lennard-Jones potential it employs. 
Thus, the SM model demonstrates a phase transition \textit{gas $\rightarrow$ isotropic liquid}, \cite{weis_levesque, pt1, pt2}, that was not observed in the DHS/DSS models. 
\\
Because of the interest in the magnetic properties of the Stockmayer model, a large part of its study has been performed with an external field applied. The introduction of an external field, however, originates a special direction in space which might have a bearing upon the behaviour of the system.\\
\\
Despite the great number of theoretical and simulation studies on the Stockmayer model, the full phase behaviour, including effects due to the entropy, has so far not been presented. We suggest a new technique of investigating based on measuring the local order in the DS models. It is worthwhile to note that the results of our calculations are in accordance with previous reports on the SM model.

Study of the SM model might also shed light upon specific features of water. Each water molecule has a high dipole value that corresponds to dipoles in DS models. In the same time the Lennard-Jones potential is frequently used in molecular-dynamical models on water, e.g. TIP3P or TIP4P, and apparently approximates fairly the intermolecular interaction between its particles. The combination of these facts brings us to the idea of considering the Stockmayer fluid as an extreme model on water.\\
\\
Another fact that speaks in favour of this thesis is that the particles of both water and SM fluid contain the feature of self-organization into complex structures, such as chains (fig. \ref{fig:chains}) or clusters of particles, though the mechanisms of such organization are different.

Let us consider a liquid of elongated molecules (e.g. rod or bar shaped), with its particles somehow ordered in space. If this order is stable, i.e. conserves for a long time period, this liquid is \textit{a liquid crystal} (LC): it flows under shear stress and at the same time has the property of an anisotropic arrangement of molecules, which is typical for crystals. Depending on the character of such order, three types of LC phases are usually distinguished: \textit{nematic}, \textit{cholesteric} and \textit{smectic}.\\
\\
In the nematic phase all molecules are approximately directed along one line called the nematic axis.
The microscopical structure of the other two is more complicated: these liquid crystals have a lamellar structure with molecules co-orientated along certain axis in each layer. The difference between the directions of the axes of neighbour layers is usually small.\\
\\

\begin{figure}[H]
\center{
\includegraphics[width=5.5cm]{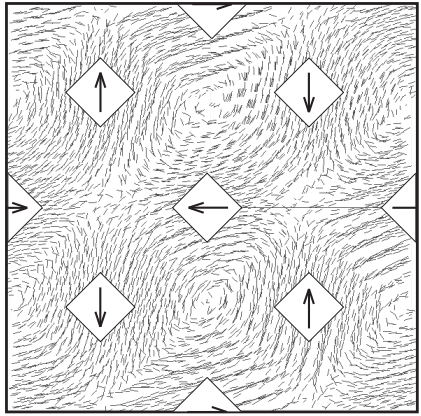}
\caption{A phase with local orientational order does not have global orientational order. Projections of dipoles on a plane are shown.}
\label{fig:local_order}
}
\end{figure}

Considering a microscopical structure of liquid crystals, we would like to pay attention to an important property of \textit{local orientational order}. In a perfect nematic phase every two molecules are almost co-directed, i.e. a global orientational order takes place. In a real liquid crystal, however, the property of co-orientation weakens with distance since the direction of the nematic axis may vary for different parts of nematic sample. \\
\\
In a substance with local orientational order the directions of two molecules at a short distance are correlated, and the correlation decreases with distance. A fluid with global orientational order is a locally orientated fluid, but not \textit{vice versa}.
\\
During our calculations we have observed the similar effect of local ordering in low temperature phases of the Stockmayer fluid. However, in this case we dealt with the orientation of dipoles instead of such of molecules. The utilization of liquid crystal terminology produces similar terms for phases of the SM model. Thus, in paper \cite{orient_order}, phases of hard and soft spheres with orientational order are called \textit{ferroelectric nematic}. The word ferroelectric underlines the dipolar meaning of the orientation of particles.\\
\\
Recent studies (as well as our calculations) confirm the presence of local orientational order in low-temperature phases of the Stockmayer fluid. The presence of local order, however, does not generally imply an existence of any order at global range, fig. \ref{fig:local_order}. 
 
Since the presence of local orientational order is a distinctive feature of low-temperature phases, we shall make an attempt to describe phase behaviour of the Stockmayer fluid from this point of view. 

\section{LOCAL ORDER PARAMETER}

According to the general theory of Landau -- de Gennes, a second order phase transition can be described by means of order parameter. In our case this is a quantity that can put apart phases with local order and without it. Thus, the following discussion is performed in framework of the Landau -- de Gennes theory. 
 
The main point of this work is our suggestion of a novel order parameter with the latter ability in the following form:
\begin{equation}
\label{eq:op}
   G = \left\langle\frac{1}{N}\sum_{0\leq i<j\leq N}\frac{R^2}{|\vec d_i||\vec d_j|}\frac{\vec d_i \cdot \vec d_j}{(\vec r_i \; - \; \vec r_j)^2}
                \right\rangle.
\end{equation}

Any pair of co-directional dipoles in SM fluid increases the order parameter just as any opposite-directed pair decreases it.
Moreover, this influence weakens rapidly with the distance increasing between the elements in such pairs. As a result, the value of G depends basically on the close-ranged orientation of dipoles in the SM fluid. In this manner the suggested order parameter detects a local order of the model and neglects any long-range disorder.\\
\\
It is worth noting that this order parameter is not the only one that takes into account the local order of particles.
The one we use is simple to count and therefore rather practical.\\
\\
It is worth noting, that in spite of the fact that the presence of local orientational order in the Stockmayer model has been reported in previous studies,  we could not find any quantitative description of this property crucial for the analysis of the phase behaviour that we performed.

\section{SIMULATION METHOD}

We performed a Monte Carlo (MC) calculation of a cubic box with Stockmayer particles in an NVT-ensemble for an interval of temperatures with fixed these basic parameters: \\
Lennard-Jones depth $\epsilon$ is 0.25 for all particles, \\
dipoles value $D$ is 5 for all particles, \\
particle radii - $R$ - are 1, \\
particles number $N$ is 2000 or 5000 (all calculations were made for both parameters). \\
\\
Since we have to perform a Monte-Carlo simulation for a large range of temperatures, there is a need for accelerating our calculations. To overcome this difficulty we apply a method that is similar to the MC annealing scheme, \cite{annealing}.
The algorithm starts from a random particles distribution at high temperature (i.e. in a gas phase). After about $1\div2 \cdot 10^8$ MC steps the calculation stops and the current system conformation becomes a start condition for the new computation with a slightly decreased temperature.
This technique allows us to reduce a number of MC steps per temperature and thus significantly reduce calculation time. \\
\\
It is also worth noting that the annealing method is used for finding the global minimum of a function of several parameters. 
Thus, we may expect to see the system we simulate in the lowest energy state at zero temperature. Such behaviour is expected from a real system as well. This fact supports this technique. The process we simulate can thus be corresponded to slow cooling of a real system.

\section{RESULTS}

We shall utilize the notation used in paper \cite{weis_levesque} for the reduced density $\rho^* = \frac{8NR^3}{V}$ with $V$ -- cell volume, $N$ -- number of particles and $R$ -- particle radius.\\
The temperature scale $T^*$ we shall use is based on the following estimates. If colloid particles have radii of order $\sim 10^3 \AA = 10^{-5} cm$ and  the magnitude of its dipoles is $\sim 10\, Debye$, we can thus roughly estimate the dipole-dipole interaction energy as 
$$
	U_{dd} = \frac{D^2}{R^3} \sim \frac{100 \cdot cm^2 \cdot \bar e^2}{(10^{-5})^3 \cdot cm^3} = 10^{-13} erg.
$$
Using this value as an estimate of the mean energy of a single particle in the fluid ($kT^* \sim U_{dd}$) we obtain for T=300K
$$
	\frac{kT}{kT^*} = \frac{1,38 \cdot 10^{-16} \cdot 300}{10^{-13}} \approx 0,41
$$
Thus, $0,41 kT^*$ corresponds to the temperature of 300K. Despite the roughness of these estimates, the order of its magnitudes indicates that the results we obtain are physically adequate. \\

\begin{figure}[H]
\center{
\includegraphics[width=8cm]{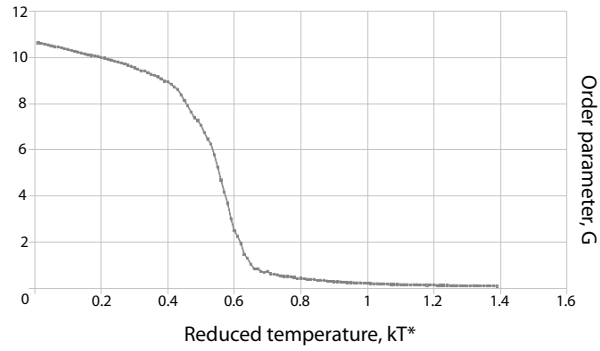}
\caption{Temperature dependence of order parameter G. Reduced density $\rho^*=0,16$, number of particles N=5000. }
\label{fig:OP}
}
\end{figure}

\begin{figure}[H]
\center{
\includegraphics[width=8cm]{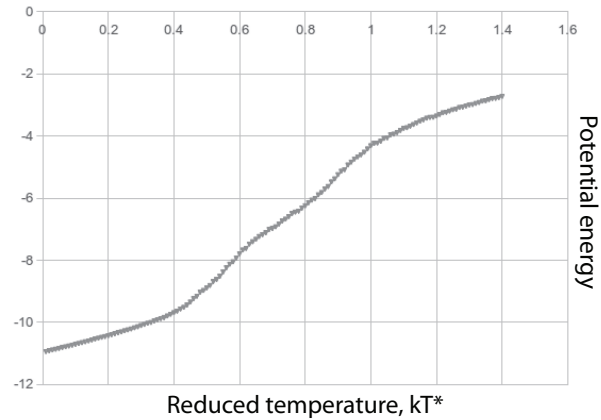}
\caption{Temperature dependence of potential energy. Reduced density $\rho^*=0,16$, number of particles N=5000.}
\label{fig:EN}
}
\end{figure}

Let us now consider a sample of the typical behaviour of the local order parameter $G$ and the potential energy of the system. The figures \ref{fig:OP} and \ref{fig:EN} contain temperature dependencies of these quantities.\\

We shall pay our attention to the following reduced temperature regions in the graph \ref{fig:OP}: \\
\\
\textit{A : 0.01 -- 0.42 :} corresponds to a phase with a local orientational order (dipoles of the SM particles are co-orientated at close range); \\
\\
\textit{B : 0.42 -- 0.65 :} corresponds to a zone of phase transition. The behaviour of the order parameter is similar to that of magnetization in the Ising model and indicates a second order transition (in Erenfest's classification); \\
\\
\textit{C : 0.65 -- 1.40 :} is the range of the locally (and globally) orientationally disordered phase.\\
\\
It is necessary to make a correction in the classification above. There is one more interesting point \textit{c`} at $kT^* \approx 0,9$. This point is hard to see on the order parameter graph \ref{fig:OP} because of the large scale, but one can notice it on the energy graph \ref{fig:EN}. At point \textit{c`} 
both graphs are bent, or speaking mathematically, have a discontinuity of first derivative (not in a precise sense however). Thus, this point corresponds to the second order \textit{gas $\rightarrow$ liquid} phase transition.\\
From this point of view, the B region (see above) corresponds to the \textit{disordered liquid $\rightarrow$ liquid with local orientational order} phase transition.\\
\\
There are two additional points in the fig. \ref{fig:OP} that arouse interest: $kT^* \approx 0.46 $ and $kT^* \approx 0.55$. These points correspond to the local maxima of the first derivative of $G$. We shall recall them in the further text.\\
\\
We would like to underline the fact that the potential energy behaviour does not provide phase information as clearly as the order parameter does.\\ 

\begin{figure}[H]
\center{
\includegraphics[width=8cm]{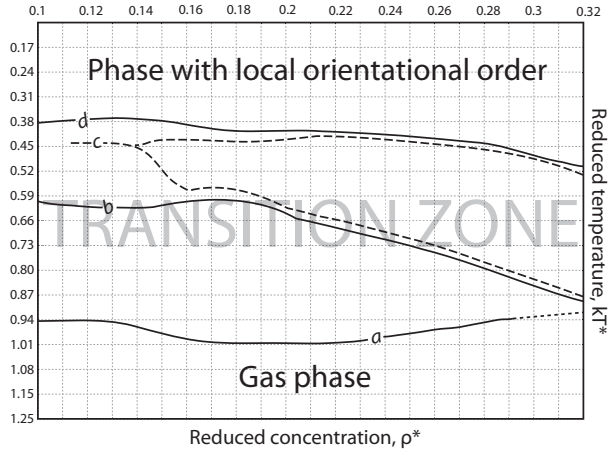}
\caption{Phase diagram of the Stockmayer fluid based on the local order parameter. Solid lines \textit{a, b, d} correspond to conformational transitions.
Dashed line \textit{c} designates local maxima of first derivative of the local order parameter G.}
\label{fig:main}
}
\end{figure}

\begin{figure}[H]
\center{
\includegraphics[width=8cm]{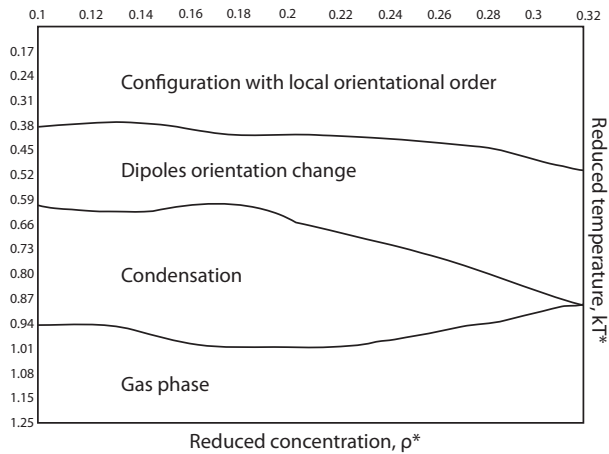}
\caption{Phase diagram of the Stockmayer liquid based on the local order parameter. Four basic phases are distinguished.}
\label{fig:phases}
}
\end{figure}

Let us now consider phase diagrams based on the behaviour of order parameter, fig. \ref{fig:main}, \ref{fig:phases}. The graph on fig. \ref{fig:OP} is a vertical section ($kT^*=const$) of the phase diagrams \ref{fig:main} and \ref{fig:phases}. The regions \textit{A}, \textit{B}, \textit{C} of the graph \ref{fig:OP} now correspond to large zones of phases, bounded by the lines \textit{d} and \textit{b} (fig. \ref{fig:main}). The point \textit{c`} correspond to the \textit{a} line. Finally, local maxima of order parameter G correspond to dashed \textit{c} line.\\
\\
Thus, we have 3 basic lines (solid lines \textit{a}, \textit{b}, \textit{d} on fig. \ref{fig:main}), that perform the division of the phase diagram on 4 parts:\\
the zone above the \textit{d} line corresponds to a phase with local orientational order;\\
in the region between the \textit{b} and \textit{d} lines there is a transition zone (dipoles become locally co-orientated);\\
in the region between the \textit{b} and \textit{a} lines the SM fluid has an isotropic liquid phase; \\
the region below the \textit{a} line corresponds to an isotropic gas phase.\\
\\

\begin{figure}[H]
\center{
\includegraphics[width=12cm]{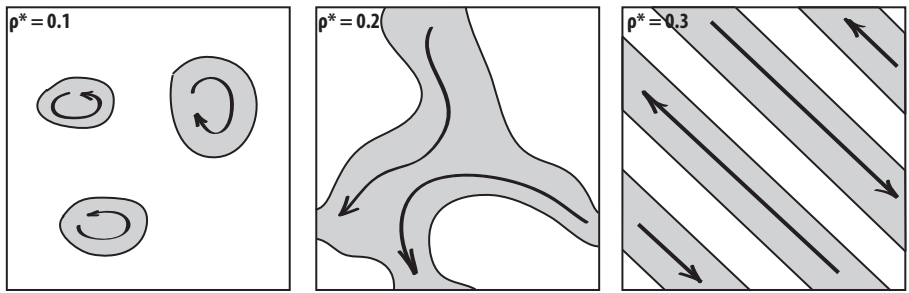}
\caption{Schematic images of three low-temperature phases. Arrows designate local directions of dipoles. The first picture corresponds to a liquid of globules with dipoles forming vortexes. The second picture corresponds to a complex network structure. The third picture represents opposite-directed nematic strands. Periodic boundary conditions are implied.}
\label{fig:confs}
}
\end{figure}

Our calculation shows that conformational organization of the Stockmayer particles depend both on temperature and the reduced density $\rho^*$. Here we perform the analysis of conformations of particles in phases at different temperatures and densities. To that end we shall consider states of the system on the last step of Monte-Carlo algorithm, calculated for different values of $kT^*$ and $\rho^*$.\\
\\
In the lowest zone (below the \textit{a} line) on figure \ref{fig:main} the fluid exhibits a uniform isotropic gas phase for all considered values of density. The region between the \textit{a} and the \textit{b} lines demonstrates an isotropic phase with density fluctuations. Based on the presence of clusters we  regard this phase as liquid.
\\
The decrease in temperature above the \textit{b} line results in the appearance of a weak local order of dipoles. At this point dipolar vortexes and regions of co-orientation of dipoles appear in dense parts of medium. The cooling of the system from the \textit{b} to the \textit{d} line contains a cascade of two-step reconstructions of the fluid: (1) particle displacement and formation of a new topological structure followed by (2) dipoles re-organization induced by new positions of particles.\\
The second part of each of these reconstructions corresponds to a local maximum of first derivative of local order parameter \textit{G}, points that we have shown as a dashed line on the fig. \ref{fig:main}. The double dashed line in the high-concentration region stands for two reconstructions we managed to distinguish in this interval of the reduced densities. \\
We feel that this cascade is an interesting effect due to entropy, and it needs to be examined in more detail. \\
Thus, during our calculation we observed a reconstruction that is reminiscent of the BKT transition in XY-model: a phase with vortexes transformed to a phase with linearly co-orientated parts in a very short range of temperature (about $0.03 kT^*$).\\
\\
The Stockmayer fluid phases at low temperatures (above the \textit{d} line) are locally orientationally ordered. In this region of the phase diagram we could distinguish three basic types of low-temperature conformations: \textit{liquid of clots}, \textit{`network` phase} and \textit{phase of opposite-directed nematic strands}, see fig. \ref{fig:confs}. It is worthwhile to note, that the existence of the latter phase has not been reported before, in spite of the vast number of publications on the topic. We were unable to determine the exact concentration boundary between these low-temperature conformations; these transitions could be gradual. This region of phase diagram is to be investigated in more detail.

\section{CONCLUSIONS}

Despite the large number of theoretical and computational studies, we are still far from understanding the phase behaviour of the Stockmayer fluid. The complexity of effects due to the entropy in the model and flaws of both theoretical and computational methods of research demand developing novel techniques in the study of the Stockmayer fluid. \\
In the current paper we suggest such a technique, which is based on a notion of local orientational order. The property of local orientational order is an important characteristic of low-temperature phases of the Stockmayer fluid. We have suggested a novel order parameter $G$ as a measure of such order. The utilization of this parameter allowed us to perform an analysis of the phase behaviour of the Stockmayer fluid from a new point of view. \\
Thus, using our technique resulted in the construction of the phase diagram (fig. \ref{fig:main}). We also managed to describe the \textit{isotropic liquid $\rightarrow$ locally ordered liquid} phase transition and find a low-temperature phase, that composed of opposite directed parallel nematic layers (fig. \ref{fig:confs}, right). The full phase diagram (for $0 \leq \rho^* \leq 1$) and analysis of conformations for different values of kT and $\rho^*$ shall be presented in our further publications.
\\
The local order we introduced provides a new point of view on systems without global order. It might be useful to discover hard-to-find inner symmetries in other soft matter systems. Specifically, we expect this method to be useful in models related to liquid crystals or water. It might also provide new results in self-assembly models.\\
\\
During our research we have also used a Monte-Carlo annealing technique that turned out to be a very effective numerical instrument for studying soft matter systems.
\\
This work was supported by RFBR, grants \#11-02-01462-a, \#10-01-00748 and by the Government Grant of the Russian Federation for support of research projects implemented by leading scientists, Lomonosov Moscow State University under the agreement No. 11.G34.31.0054.

\end{document}